\documentclass[a4paper]{article}

\usepackage{INTERSPEECH2022}
\usepackage{multirow}
\usepackage{todonotes}
\usepackage{url}

\title{Automated Evaluation of Standardized Dementia Screening Tests}
\name{F. Braun$^1$, M. Förstel$^1$, B. Oppermann$^1$,  A. Erzigkeit$^2$, T. Hillemacher$^3$, H. Lehfeld$^3$, K. Riedhammer$^1$}
\address{
  $^1$Technische Hochschule Nürnberg Georg Simon Ohm, Germany\\
  $^2$Psychiatrische Klinik und Psychotherapie, Universitätsklinikum Erlangen, Germany\\
  $^3$Klinik für Psychiatrie und Psychotherapie, Universitätsklinik der Paracelsus Medizinischen Privatuniversität, Klinikum Nürnberg, Germany
}
\email{korbinian.riedhammer@th-nuernberg.de,hartmut.lehfeld@klinikum-nuernberg.de}
\begin{document}

\maketitle
\begin{abstract}
For dementia screening and monitoring, standardized tests play a key role in clinical routine since they aim at minimizing subjectivity by measuring performance on a variety of cognitive tasks.
%
In this paper, we report on a study that consists of a semi-standardized history taking followed by two standardized neuropsychological tests, namely the SKT and the CERAD-NB.
The tests include basic tasks such as naming objects, learning word lists, but also widely used tools such as the MMSE.
Most of the tasks are performed verbally and should thus be suitable for automated scoring based on transcripts.
For the first batch of 30 patients, we analyze the correlation between expert manual evaluations and automatic evaluations based on manual and automatic transcriptions.
For both SKT and CERAD-NB, we observe high to perfect correlations using manual transcripts; for certain tasks with lower correlation, the automatic scoring is stricter than the human reference since it is limited to the audio.
Using automatic transcriptions, correlations drop as expected and are related to recognition accuracy; however, we still observe high correlations of up to 0.98 (SKT) and 0.85 (CERAD-NB).
We show that using word alternatives helps to mitigate recognition errors and subsequently improves correlation with expert scores.

\end{abstract}

\section{Introduction}
Dementia is characterized by a loss or decline of function; in addition to memory impairments, patients exhibit one or more of aphasia, apraxia, agnosia or impairments of executive function.
These symptoms can relate to different psychological (e.g., trauma) or neurological conditions (e.g., Alzheimer's).
Early diagnostic clarification with the resulting possibility of a rapid start of treatment is key to delaying the progression of dementia and thus achieving a gain in quality of life for the patient and their family caregivers.
Dementia screening and monitoring enable early detection, classification and tracking of cognitive decline.
To that end, standardized tests play a key role in clinical routine since they aim at minimizing subjectivity by measuring performance on a variety of cognitive tasks.
Tests typically target both short- and long-term memory and cover tasks such as naming, memorizing, counting and recalling objects, or general situational awareness.
Such tests are administered by trained physicians or psychologists who spend about 15--45 minutes with the patient.
With appointment wait times of up to 6 months and longer, automated tests could help to closely monitor patients and prioritize urgent cases for in-person appointments.

In this paper, we report on a study conducted at the Memory Clinic of the Department of Psychiatry and Psychotherapy, Nuremberg Hospital, Germany, that consists of a semi-standardized history taking followed by two standardized neuropsychological tests, namely the Syndrom-Kurz-Test (SKT, translates to ``Syndrome Short Test'') and the Consortium to Establish a Register for Alzheimer's Disease Neuropsychological Battery (CERAD-NB).
With tasks such as naming objects, learning word lists, but also widely used tools such as the Mini-Mental State Examination (MMSE), these tests comprise a broad inventory for determining the cognitive state.
We recorded the conversations between patients and therapists during the screening sessions to obtain a dataset containing spontaneous speech from the history taking and elicited speech from performing the tests. 
Since most of the tasks are done verbally, they should be suitable for automated scoring based on transcripts.
For a first batch of 30 patients, we analyze the correlation between expert manual evaluations and automatic evaluations based on manual and automatic transcriptions.
We test how the use of word alternatives helps to reduce recognition errors and subsequently improve correlation with expert ratings.
Furthermore, we investigate the effect of eliminating outliers that have a particularly poor recognition performance.

\section{Related Work}
In addition to medical examinations (e.g., brain imaging), combinations of medical and psychological history taking, cognitive testing, and rating scales are the gold standard for dementia screening in clinical or research settings \cite{cooper05}.
The widely used Mini-Mental State Examination (MMSE), the Clock Drawing Test (CDT), the Mini-Cog test, the SKT (German) \cite{erzigkeit15}, and many other cognitive scales have gained acceptance since they are brief while still showing good sensitivity and specificity \cite{sheehan12}.
Neuropsychological test batteries such as the Boston Diagnostic Aphasia Exam (BDAE, \cite{borod80}) and the CERAD-NB \cite{morris89} evaluate various perceptual modalities (auditory, visual, gestural), processing functions (comprehension, analysis, problem-solving) and response modalities (writing, articulation, manipulation).
They include common sub-tests such as the Cookie Theft Picture Description (CTPD), the Boston Naming Test (BNT), and the Verbal Fluency Test (VFT).

The automation of dementia screening based on speech is an area of high interest; it was previously covered by the ADReSS and ADReSSo Challenges \cite{adress20,adresso21}.
Previous work shows strong evidence for the effectiveness of speech-based screening in dementia patients, even at early stages, and focuses primarily on the publicly available DementiaBank \cite{becker1994dementiabank}.
\cite{adress20,fraser16,alhameed16,koenig15,orimaye17} obtained convincing results on spontaneous speech of the CTPD from the BDAE.
Free recall tasks of visual material, such as the CTPD, have the advantage that spontaneous speech is elicited and retrieved selectively, making it more contained and thus easier to process.
The same is true for elicited speech based on free recall tasks from moving images, such as short films \cite{vincze22}.
Most work uses either fluency \cite{koenig18,frankenberg21} or deep speech markers \cite{adresso21} for classification, as these show high variance in distinguishing patients with cognitive impairment from healthy controls.

There also is work on direct automated assessment in dementia screening that focus on drawing tests such as figure copying \cite{webb21} and CDT \cite{prange19}, or on VFT \cite{troeger18,kwon21,kim19}.
However, we are not aware of any work that has performed automatic evaluation of the entire SKT and CERAD-NB.

\section{Standardized Tests}
\subsection{Syndrom-Kurz-Test (SKT)}
The Syndrom-Kurz-Test (SKT)  is a short performance test for the detection of disorders of memory and attention \cite{erzigkeit77}.
It consists of a total of nine sub-tests: three for recording memory performance (reproducing objects immediately, reproducing objects after distraction, recognizing objects) and six for recording attention in terms of information processing speed (naming objects, reading numbers, ordering numbers, reordering numbers, counting symbols, interference test) \cite{erzigkeit15}.
In the attention sub-tests, the time taken to complete the task is measured, while in the memory sub-tests, the number of missing objects is counted and confabulations\footnote{Confabulations are mentions that did not belong to the objects, and are a typical sign of dementia, as those affected try to fill in their memory gaps.
They are noted as an indication, but are not included in the test score.} are noted.
All sub-tests have a time limit of 60 seconds, so that the duration of the entire test is 10 to 15 minutes.
To prevent a learning effect in dementia monitoring, there are five forms of the SKT (A--E) that map the same tasks with different content (pictures, numbers, symbols).
For the automatic evaluation based on transcripts, only those sub-tests are considered that are done verbally; we therefore exclude the tasks for ordering (4) and reordering numbers (5) that require actions instead of speech.
The descriptions of the sub-tests are based on the official SKT manual \cite{erzigkeit15}.

\textbf{SKT 1 (naming objects):}
Twelve pictures of everyday objects are to be named. 
The subject is informed that the naming speed and memorization is important since the objects are to be reproduced from memory immediately afterwards.

\textbf{SKT 2 (reproducing objects immediately):}
The named objects from sub-test 1 are to be reproduced from memory. 
This is followed by a short learning phase: The template is shown to the subject again for five seconds, with the request to memorize the objects, since they will be asked about them again later.

\textbf{SKT 3 (reading numbers):}
Ten two-digit numbers on tokens are to be read aloud as quickly as possible in the direction of reading.

\textbf{SKT 6 (counting symbols):}
Processing a perceptual task quickly and as carefully as possible. 
The subject must recognize and correctly count a target symbol (e.g., a square) from a series of distractors.
Whether the number is correct or not is noted, but does not enter into the test score. 
The aim is to measure the processing time for scanning the entire page.

\textbf{SKT 7 (interference test):}
A sequence consisting of two repeating letters (e.g., ``A'' and ``B'') is to be read as quickly and accurately as possible.
The particular challenge is that the subject has to read one letter but say the other (i.e., read ``A'' but say ``B'' and vice versa); this measures ``disposition rigidity'' according to Cattell \cite{cattell49}.

\textbf{SKT 8 (reproducing objects after distraction):}
The subject is asked to recollect the objects shown in sub-test 1.
Here, the indirect reproduction performance is tested by the free recall of the objects.

\textbf{SKT 9 (recognizing objects):}
Measures the stable memory function ``recognition.''
The 12 objects shown in sub-test 1 are to be named from 48 displayed objects; speaking is required, pointing is invalid. 

\subsection{CERAD Neuropsychological Battery (CERAD-NB)}
The CERAD assessment instruments comprise five different domains, one of which is the Neuropsychological Assessment of Alzheimer's related dementia (AD). 
The neuropsychological test battery \cite{morris89} developed in this context (CERAD-NB) measures brain performance from those functional domains in which specific cognitive deficits can be observed in AD, namely memory, language, praxia, and orientation. 
The standard CERAD-NB includes a verbal fluency test (listing animals) \cite{isaacs73}, the BNT (confrontation naming) \cite{kaplan78}, the MMSE \cite{mmse75}, word list learning, recall and recognition \cite{atkinson71}, as well as figures drawing and recalling \cite{rosen84}.
Again, we exclude those tasks that are done without speech, which in this case are figures drawing (5) and recalling (8).
The descriptions of the sub-tests are based on the official CERAD-Plus Online Manual \cite{cerad15}.

\textbf{CERAD-NB 1 (verbal fluency test):}
The subject is asked to list as many different animals as possible for one minute; the number of correctly named animals is scored.
This task is used to examine the speed and ease of verbal production ability, semantic memory, linguistic ability, executive functions, and cognitive flexibility. 

\textbf{CERAD-NB 2 (Boston Naming Test):}
A confrontation naming test, where the subject is asked to name objects, which are represented in the form of line drawings. 
The names of the objects are differentiated according to the frequency of their occurrence in the (American) language: frequent, medium, rare. 
The number of objects correctly named spontaneously is counted.
This task measures linguistic ability, visual perception, and naming, respectively word finding.
The original version of the BNT includes 60 line drawings, while the version in the CERAD-NB is a 15-drawing short form; it is adapted to the German language.

\textbf{CERAD-NB 3 (Mini-Mental State Examination):}
A widely used screening instrument to assess basic cognitive abilities, focusing on temporal and spatial orientation, linguistic performance (especially linguistic memory function), concentration, memory, recall, and constructive practice; the maximum score is 30 points. 
In contrast to the original MMSE, the CERAD-NB replaced the serial subtraction task ``100-7'' with spelling the word ``world'' (resp. ``Preis'' in the German version) backwards. 

\textbf{CERAD-NB 4 (word list learning):}
The subject has to read 10 printed words aloud in succession and is then asked to recall them freely from memory. 
In two subsequent runs, the same words are shown again in a different order and are then each to be reproduced freely again. 
The maximum number of correct answers over all three runs is 30 (word list learning total).
This task measures the ability to learn new, non-associated verbal information. 
The ``Word List Intrusions'' are calculated, which are the equivalent of confabulations in SKT and contain words that were not included in the word list and were misnamed during recall. 
The sum of the intrusions from the three learning passes and from the delayed recall (CERAD-NB 6) is calculated. 

\textbf{CERAD-NB 6 (word list recall):}
After a time delay and distraction by CERAD-NB 5 (drawing figures), the subject is asked to recall the 10 words memorized in Task 4; the maximum number of correct answers is 10.
This tests verbal episodic memory, that is, whether a person can retain newly learned verbal information over a period of several minutes. 
As an additional variable the ``Word List Savings'' $S$, which is a value for verbal retentiveness, is calculated as follows \cite{cerad15}:
$$S = \frac{\text{Word List Recall}}{\text{Word List Learning Pass 3}} \times 100$$

\textbf{CERAD-NB 7 (word list recognition):}
With the word list recognition test (discriminability) \cite{mohs86} the subject is presented with 20 words, including the 10 words from task 4, mixed with 10 distractions.
For each word, the person must decide whether or not it belongs to the previously memorized words.
The discriminability $D$ (in \%) is calculated according to the following formula \cite{mohs86}:
$$D = \left\{ \frac{\left ( 10 - \text{Hits} \right ) + \left ( 10 - \text{Correct Rejections} \right )}{20}\right\} \times 100$$
This tests whether the subject can benefit from facilitated recall conditions, and thus helps to distinguish whether deficits are primarily in recall or in memory.

\section{Data}
All dementia screenings were carried out at the Memory Clinic (``Gedächtnissprechstunde'') of the Department of Psychiatry and Psychotherapy, Nuremberg Hospital, Germany.\footnote{Research approved by the Ethics Committee of the Klinikum Nürnberg under File No. IRB-2021-021; each subject gave informed consent prior to recording.}
To date, a total of 103 German-speaking subjects participated in the study, of which 30 were fully transcribed and included at time of writing. 
The included subjects range in age from 59 to 88 years ($\mu = 73.5 \pm 8.4$) and are balanced in sex (15f, 15m).
Their medical diagnoses range from mild cognitive impairment (MCI) to mild and moderate dementia.

All participants underwent the three-part screening procedure:
history taking with questions on memory, attention, daily living competence, orientation, mood, sleep, appetite, physical activity, and medication; SKT and CERAD-NB tests; two questionnaires for self-assessment of mood (GDS-K: Geriatric Depression Scale Short Form) and daily living competence (B-ADL: Bayer-Activities of Daily Living Scale).

Data includes labels for SKT and CERAD-NB sub- and total scores, both as raw and normalized values, as well as coded medical and psychological diagnoses.
Metadata includes sex, age, smoker/non-smoker, medication (antidementives, antidepressants, analgesics), GDS-K, B-ADL (self assessment and external assessment), NPI (Neuropsychiatric Inventory, external assessment), IQ-range ($<$90, 90--110, $>$110), and years of education.
Furthermore, we labeled the data with start and end times for each of the sub-tests.

Recording was performed with a Zoom H2n Handy Recorder in XY stereophonic mode, positioned between the patient and the therapist in such a way that level differences between the left (therapist) and right (patient) channels could be used to separate the speakers.
The audio samples were recorded in 16-bit stereo wav format with a sampling rate of 48 kHz and later converted to uncompressed PCM mono wav format with a sampling rate of 16 kHz.
The audio recordings include about 23.5 hours of speech, transcribed by a professional transcription service according to the transcription rules of Dresing \& Pehl.
Both therapists and patients reported that they were not affected by the presence of the device.
Due to the Corona pandemic, therapists and patients wore surgical or KN95 masks that affect the speech signal according to \cite{nguyen21}.
The speech of some subjects exhibits strong forms of local dialects.

\section{Experiments}

\subsection{Manual Test Scores}
\label{Sec:manual_scores}
Tests were conducted in-person and manually evaluated by a psychotherapist. 
In the case of the SKT, the raw test scores, that is, the number of words not remembered or the processing time, are converted into norm values, which are normalized according to age and IQ (below average, average, above average); a total score of 0--27 indicates the degree of cognitive impairment in the case of homogeneous sub-score profiles according to \cite{erzigkeit15}.
The SKT test evaluation using the ``old'' SKT norms (c.f. Table 11 in \cite{erzigkeit15}) differentiates between the degree of no cognitive impairment (0--4), mild cognitive impairment (5--8), as well as mild (9--13), moderate (14--18), severe (19--23), and very severe (24--27) dementia syndrome.
SKT memory and attention sub-scores provide a rating of no to very severe memory or attention impairment.

In the case of CERAD-NB, z-values are calculated based on a complex normalization formula \cite{berres00}, which includes the influence of age, education level, and gender.




\subsection{Automated Scoring}
As described previously, we focus on the (sub-)tests that can be evaluated from an audio recording and respective transcript.
In addition to the manual transcripts, automatic speech recognition (ASR) was used to obtain regular 1-best transcriptions as well as transcriptions with up to 5 word alternatives (sometimes referred to as ``sausages'').
We used the Mod9 ASR Engine\footnote{\url{https://mod9.io}} with the \verb|word-alternatives| option and the \verb|tuda_swc_mailabs_cv_voc683k| acoustic and language models provided by \cite{milde2018german}.
Word recognition rate (accuracy) ranges from about 10\% to 50\% on 1-best, improving to about 20\% to 60\% using the word alternatives.\footnote{``Percent Correct,'' using NIST SCTK, \url{https://github.com/usnistgov/SCTK}}
To further analyze the impact of transcription errors on correlation, we also consider a subset of 21 subjects that show an accuracy above 20\% (``Top-21'').


We compute correlations between the expert and automatic test scores based on manual and automatic transcripts.
The following sections detail the automatic scoring for each of the (sub-)tests.

\subsection{SKT}


\begin{table}
 \setlength{\tabcolsep}{0.41\tabcolsep} 
 \centering
 \caption{\label{Tab:SKT_evaluation}Automated SKT scoring on manual transcriptions (Trans.) and automatic speech recognition with (ASR-5) and without (ASR-1) the top five word alternatives.
 Column Top-21 refers to top 21 speakers and ASR-5.}
 \begin{tabular}{ c|c|ccc|c } 
  \toprule
  \textbf{ID} & \textbf{Test/Task} & \textbf{Trans.} & \textbf{ASR-1} & \textbf{ASR-5} & \textbf{Top-21}\\
  \midrule
  1 & naming objects & 0.89 & 0.70 & 0.81 & 0.89 \\ 
  2 & reproducing objects & 1.00 & 0.58 & 0.71 & 0.83 \\
  3 & reading numbers & 0.94 & 0.85 & 0.86 & 0.94 \\
  6 & counting symbols & 0.90 & 0.59 & 0.58 & 0.54 \\
  7 & interference test & 0.99 & 0.97 & 0.98 & 0.99 \\
  8 & naming after distraction & 1.00 & 0.75 & 0.90 & 0.97 \\
  9 & recognizing objects & 0.89 & 0.50 & 0.55 & 0.68 \\ \hline
  - & attention score & 0.92 & 0.84 & 0.82 & 0.85 \\
  - & memory score & 0.98 & 0.62 & 0.78 & 0.93\\
  \bottomrule
  - & total score & 0.97 & 0.81 & 0.89 & 0.94 \\ \hline
 \end{tabular}
\end{table}

The test scoring is done for each task group separately.
In the first group, consisting of tasks 1 (naming objects), 2 (reproducing objects immediately), 8 (reproducing objects after distraction) and 9 (recognizing objects), we count the number of objects recognized and store them with their timestamps. 
For task 1, we then calculate the duration using the start time of the first word and the end time of the last word, which represents the score. 
For the results of tasks 2, 8, and 9, the number of missing objects is calculated based on the number of recognized objects. 
For task 3 (reading numbers), the timestamp of the first and last two-digit number is used. 
The score of task 6 (counting symbols) is calculated by the individual timestamps from the end of the task and the last two-digit number named.
From the timestamps of the first and last letter mentioned, we calculate the duration that gives the score for task 7 (interference test).

\tablename~\ref{Tab:SKT_evaluation} shows the correlations of the expert scoring with the automatic scoring based on manual and automatic transcriptions.
Since correlations for automatic scoring based on manual transcripts are not always close to perfect, we did some error analysis to improve our algorithms:
For task 1, we encountered the issue that the objects were named multiple times by some patients, following further instruction by the examiner.
For task 3, transcript normalization for numbers can be an issue.
For task 6, subjects typically count aloud; the examiner sometimes assists.
For task 9, some patients only pointed to the individual objects instead of naming the objects.
In addition, overall accuracy was affected by manual creation of timestamps based on forced alignment and transcription errors. 

Using automatic transcripts for the automated scoring leads to lower correlations, which is expected due to recognition errors.
To achieve higher word recognition, we use the top five word alternatives generated by the ASR system.
As noted in the data description, all speakers wore face masks and partially exhibited local dialect.
Excluding those subjects with recognition accuracy below 20\% restores correlations similar to those based on human transcription.

\subsection{CERAD-NB}

\begin{table}
 \centering
 \caption{\label{Tab:CERAD_evaluation}Automated CERAD scoring on manual transcriptions (Trans.) and automatic speech recognition with (ASR-5) and without (ASR-1) the top five word alternatives. Column Top-21 refers to top 21 speakers and ASR-5.}
 \setlength{\tabcolsep}{0.58\tabcolsep}
 \begin{tabular}{ c|c|ccc|c } 
  \toprule
  \textbf{ID} & \textbf{Test} & \textbf{Trans.} & \textbf{ASR-1} & \textbf{ASR-5} & \textbf{Top-21} \\
  \midrule
  1 & verbal fluency test & 0.98 & 0.82 & 0.85 & 0.91 \\ 
  2 & Boston Naming Test & 0.70 & 0.14 & 0.24 & 0.47 \\
  3 & MMSE & 0.71 & 0.07 & 0.35 & 0.52 \\
  4 & word list learning & 0.94 & 0.62 & 0.70 & 0.75 \\
  6 & word list recall & 0.99 & 0.68 & 0.81 & 0.78 \\
  \bottomrule
  - & total & 0.71 & 0.37 & 0.49 & 0.61 \\
  \hline
 \end{tabular}
\end{table}

We use several task-specific methods for the automatic scoring of the CERAD-NB.
To calculate the score of the CERAD-NB task 1 (verbal fluency test) we use spaCy \cite{Honnibal_spaCy_Industrial-strength_Natural_2020} to normalize all named words to their base form, and then use GermaNet \cite{hamp-feldweg-1997-germanet, henrich-hinrichs-2010-gernedit} to filter all words that have a hyponymic relationship to ``animal.'' 
After filtering out duplicates, the word count represents the score.
The CERAD-NB tasks 2 (Boston Naming Test), 4 (word list learning), 6 (word list recognition), and parts of 3 (MMSE) can be scored by counting the required words, allowing for some hand-picked alternatives (e.g., dromedary instead of camel).
We use a weighted Levenshtein distance to evaluate the CERAD-NB 3 subtasks that require the patient to repeat a word and then spell it backwards.
To account for filler words and different spelling behavior (e.g., using names to spell letters like ``Alice'' for ``A''), we assign no cost to deletions.
Since not all subtasks of CERAD-NB 3 are suitable for automatic scoring, we scale the scorable subtasks to account for the missing ones.
We exclude CERAD-NB task 7 (word list recognition) from our tests because of the highly heterogeneous response patterns of patients.
When scoring based on automatic transcripts, we perform tests and report correlations with (ASR-5) and without (ASR-1) the top five word alternatives.
We also perform tests in which we omit the patients with the lowest ASR word recognition (i.e., the patients with whom the speech recognizer had the most problems).
We calculate the total score according to \cite{Chandler102} which has been proven to also work well on the German version of the CERAD-NB \cite{Ehrensperger2010-il}. 

\tablename~\ref{Tab:CERAD_evaluation} shows the correlation between the expert scores and the automatically scored tasks.
CERAD-NB tasks 1, 4, and 6 achieve near perfect correlation on the manual transcripts.
The CERAD-NB tasks 2 and 3 show a lower but still high correlation.
We assume that the lower correlation is due to low score variance ($\mu = 14.2 \pm1.2$) in CERAD-NB 2 and the non-scorable subtasks of CERAD-NB 3.
Small errors also arise from non-perfect time labels on the manual transcripts, sometimes resulting in small overlaps between tests.
This is especially a problem in CERAD-NB task 4, which requires patients to recall a list of words immediately after reading them.
Unsurprisingly, ASR scores are lower overall, with better results when word alternatives are included (WR-5) and patients with low word recognition are ignored (Top-21).

\section{Summary and Outlook}
In a first study, we recorded 30 patients who underwent a 3-stage dementia screening procedure.
We successfully automated the speech-based parts of the neuropsychological tests SKT and CERAD-NB.
We demonstrated high to perfect correlations using manual transcripts.
With automated transcripts, we also partially obtained high correlations of 0.98 (SKT) and 0.85 (CERAD-NB).
By including word alternatives and excluding speakers with very low recognition accuracy, we were able to significantly increase correlations; the low accuracy but high correlation suggests that errors mostly affect non-critical words.
The error analysis showed that task-specific modifications of the speech recognizer will lead to substantial improvements in scoring; in particular, tasks that include numbers, counting, or enumerations are typically ill-formed w.r.t. general-purpose language models.
Furthermore, adapting the acoustic models to elderly speech as well as local accent and dialect will help to increase overall recognition accuracy.

In this paper, we showed preliminary results from an ongoing study that focused on automating two standardized tests.
The history taking interview is currently not considered in the evaluation, but related work shows that dementia classification based on spontaneous speech is targetable.
Therefore, we plan to predict scores and diagnoses based on acoustics and text from the interview to become independent of strict test protocols.

\bibliographystyle{IEEEtran}
\bibliography{refs}

\end{document}